# An Optimization Model for Outlier Detection in Categorical Data

Zengyou He, Xiaofei Xu, Shengchun Deng

*Department of Computer Science and Engineering Harbin Institute of Technology,*

*92 West Dazhi Street, P.O Box 315, P. R. China, 150001*

zengyouhe@yahoo.com, {xiaofei, dsc}@hit.edu.cn

**Abstract** The task of outlier detection is to find small groups of data objects that are exceptional when compared with rest large amount of data. Detection of such outliers is important for many applications such as fraud detection and customer migration. Most existing methods are designed for numeric data. They will encounter problems with real-life applications that contain categorical data. In this paper, we formally define the problem of outlier detection in categorical data as an optimization problem from a global viewpoint. Moreover, we present a local-search heuristic based algorithm for efficiently finding feasible solutions. Experimental results on real datasets and large synthetic datasets demonstrate the superiority of our model and algorithm.

**Keywords** Outlier, Optimization, Local Search, Entropy, Data Mining.

## 1. Introduction

In contrast to traditional data mining task that aims to find the general pattern applicable to the majority of data, outlier detection targets the finding of the rare data whose behavior is very exceptional when compared with rest large amount of data. Studying the extraordinary behavior of outliers helps uncovering the valuable knowledge hidden behind them and aiding the decision makers to make profit or improve the service quality. Thus, mining for outliers is an important data mining research with numerous applications, including credit card fraud detection, discovery of criminal activities in electronic commerce, weather prediction, and marketing.

A well-quoted definition of outliers is firstly given by Hawkins [1]. This definition states, "An outlier is an observation that deviates so much from other observations as to arouse suspicion that it was generated by a different mechanism . With increasing awareness on outlier detection in data mining literature, more concrete meanings of outliers are defined for solving problems in specific domains [3-37]. However, conventional approaches do not handle categorical data in a satisfactory manner, and most existing techniques lack for a solid theoretical foundation or assume underlying distributions that are not well suited for exploratory data mining applications. To fulfill this void, an optimization model is explored in this paper for mining outliers.

From a systematic viewpoint, a dataset that contains many outliers have a great amount of mess. In other words, removing outliers from a data set will result in a dataset that is less "disordered . Based on this observation, the problem of outlier mining could be defined informally as an optimization problem as follows: finding a small subset of target dataset such that the degree of disorder of the resultant dataset after the removal of this subset is minimized.

In our optimization model, we first have to resolve the issue of what we mean by the "the degree of disorder of a dataset . In other words, we have to make our objective function clear. Entropy in information theory is a good choice for measuring the "the degree of disorder of a dataset . Hence, we will aim to minimize the expected entropy of the resultant dataset in our problem.

Consequently, we have to resolve the issue of what we mean by "a small subset of target dataset . Since it is very common in the real applications to report top-$k$ outliers to end users, we set the size of this set to be $k$. That is, we aim to find $k$ outliers from the original dataset, where $k$ is the expected number of outliers in the data set.

So far, the optimization problem could be described in a more concise manner as follows: finding a subset of $k$ objects such that the expected entropy of the resultant dataset after the removal of this subset is minimized.

In the above optimization problem, an exhaustive search through all possible solutions with $k$ outliers for the one with the minimum objective value is costly since for $n$ objects and $k$ outliers there are $\binom{n}{k}$ possible solutions. A variety of well known greedy search techniques, including simulated annealing and genetic algorithms, can be tried to find a reasonable solution. We have not investigated such approaches in detail since we expect the outlier-mining problem to be mostly applied large datasets, so computationally expensive approaches become unattractive. However, to get a feel for the quality-time tradeoffs involved, we devised and studied the greedy optimization scheme that uses local-search heuristic to efficiently find feasible solutions. Experimental results on real datasets and large synthetic datasets demonstrate the superiority of our model and algorithm.

The organization of this paper is as follows. First, we present related work in Section 2. Problem formulation is provided in Section 3 and the local-search heuristic based algorithm is introduced in Section 4. The Empirical studies are provided in Section 5 and a section of concluding remarks follows.

## 2. Related Work

Previous researches on outlier detection broadly fall into the following categories.

*Distribution based* methods are in the first category, which are previously conducted by the statistics community [1,5,6]. They deploy some standard distribution model (e.g., normal) and flag as outliers those points that deviate from the model. Yamanishi et al. [7] used a Gaussian mixture model to present the normal behaviors and each datum is given a score based on changes in the model. This approach was combined with a supervised-based learning approach to obtain general patterns for outlier in [8].

*Depth-based* is the second category for outlier mining in statistics [9,10]. Based on some definition of depth, data objects are organized in convex hull layers in data space according to peeling depth, and outliers are identified as data objects with shallow depth values.

*Deviation-based* techniques identify outliers by inspecting the characteristics of objects and consider an object that deviates these features as an outlier [11].

*Distance based* method was originally proposed by Knorr and Ng [12-15]. This notion is further extended based on the distance of a point from its $k^{th}$ nearest neighbor [16]. Alternatively, the outlier factor of each data point is computed as the sum of distances from its *k* nearest neighbors in [17]. Bay and Schwabacher [18] present an algorithm with near linear time for mining distance-based outlier detection.

*Density based* This was proposed by Breunig et al. [19]. It relies on the local outlier factor (*LOF*) of each point, which depends on the local density of its neighborhood. Tang el at [20] introduces a connectivity-based outlier factor (*COF*) scheme that improves the effectiveness of *LOF* scheme. Three enhancement schemes over *LOF* are introduced in [21]. An effective algorithm for mining local outliers is proposed in [22]. The *LOCI* method [23] and low-density regularity method [24] further extended the density-based approach [19].

*Clustering-based* outlier detection techniques regarded *small* clusters as outliers [25] or identified outliers by removing clusters from the original dataset [26]. Ref. [27] proposed the concept of *cluster-based local outlier*.

*Sub Space based*. Aggarwal and Yu [3] discussed a new technique for outlier detection, which finds outliers by observing the density distribution of projections from the data. A frequent pattern based outlier detection method is proposed in [4], which aims at utilizing frequent patterns in different subspaces to define outliers in high dimensional space. Wei et al. [28] introduces an outlier mining method based on a hypergraph model to detect outliers in categorical dataset.

*Support vector based*. Support vector novelty detector (*SVND*) was recently developed. The first *SVND* is proposed by Tax and Duin [29]. Another alternative *SVND* is proposed by Scholkopf et al. [30]. Cao et al. [31] further extended the *SVND* method. Petrovskiy [32] combine kernel methods and fuzzy clustering methods.

*Neutral network based*. The replicator neutral network (*RNN*) is employed to detect outliers by Harkins et al. [33,34].

In addition, the class outlier detection problem is considered in [35-37].

## 3. Background and Problem Formulation

In this section, we present the background of entropy and outliers and formulate the problem.

### 3.1 Entropy

Entropy is the measure of information and uncertainty of a random variable [2]. Formally, if $X$ is a random variable, and $S(X)$ the set of values that $X$ can take, and $p(x)$ the probability function of X, the entropy $E(X)$ is defined as shown in Equation (1).

$$E(X) = -\sum_{x \in S(X)} p(x) log(p(x)) \quad (1)$$

The entropy of a multivariable vector $\hat{x} = \{X_1, ..., X_m\}$ can be computed as shown in Equation (2).

$$E(\hat{x}) = -\sum_{x_1 \in S(X_1)} \cdots \sum_{x_m \in S(X_m)} p(x_1, ..., x_m) log(p(x_1, ..., x_m)) \quad (2)$$

### 3.2 Problem Formulation

The problem we are trying to solve can be formulated as follows. Given a dataset $D$ of $n$ points $\hat{p}_1, \ldots, \hat{p}_n$, where each point is a multidimensional vector of $m$ categorical attributes, i.e., $\hat{p}_i = (p_i^1, \ldots, p_i^m)$, and given a integer $k$, we would like to find a subset $O \subseteq D$ with size $k$, in such a way that we minimize the entropy of $D - O$. That is,

$$\min_{O \subseteq D} E(D - O) \tag{3}$$

subject to $|O| = k$

In this problem, we need to compute the entropy of a set of records using Equation (2). To make computation more efficient, we make a simplification in the computation of entropy of a set of records. We assume the independences of the record, transforming Equation (2) into Equation (4). That is, the joint probability of combined attribute values becomes the product of the probabilities of each attribute, and hence the entropy can be computed as the sum of entropies of the attributes.

$$E(\hat{x}) = -\sum_{x_1 \in S(X_1)} \cdots \sum_{x_m \in S(X_m)} p(x_1, \ldots, x_m) \log(p(x_1, \ldots, x_m))$$

$$= E(X_1) + E(X_2) + \ldots + E(X_n) \tag{4}$$

## 4. Local Search Algorithm

In this section, we present a local-search heuristic based algorithm, denoted by LSA, which is effective and efficient on identifying outliers.

### 4.1 Overview

The LSA algorithm takes the number of desired outliers (supposed to be $k$) as input and iteratively improves the value of object function. Initially, we randomly select $k$ points and label them as outliers. In the iteration process, for each point labeled as non-outlier, its label is exchanged with each of the $k$ outliers and the entropy objective is re-evaluated. If the entropy decreases, the point's non outlier label is exchanged with the outlier label of the point that achieved the best new value and the algorithm proceeds to the next object. When all non outlier points have been checked for possible improvements, a sweep is completed. If at least one label was changed in a sweep, we initiate a new sweep. The algorithm terminates when a full sweep does not change any labels, thereby indicating that a local optimum is reached.

### 4.2 Data Structure

Given a dataset $D$ of $n$ points $\hat{p}_1, \ldots, \hat{p}_n$, where each point is a multidimensional vector of $m$ categorical attributes, we need $m$ corresponding hash tables as our basic data structure. Each hash table has attribute values as keys and the frequencies of attribute values as referred values. Thus, in $O(1)$ expected time, we can determine the frequency of an attribute value in corresponding hash table.

### 4.3 The Algorithm

Fig.1 shows the LSA algorithm. The collection of records is stored in a file on the disk and we read each record $t$ in sequence.

```
Algorithm LSA
Input:    D    // the categorical database
          k    // the number of desired outliers
Output:   k identified outliers

/* Phase 1-initialization */
01  Begin
02      foreach record t in D
03          counter++
04          if counter<=k then
05              label t as an outlier with flag "1
06          else
07              update hash tables using t
08              label t as a non-outlier with flag "0

/* Phase 2-Iteration */
09      Repeat
10          not_moved =true
11          while not end of the database do
12              read next record t which is labeled "0     //non-outlier
13              foreach record o in current k outliers
14                  exchanging the label of t with that of o and evaluating the change of entropy
15              if maximal decrease on entropy is achieved by record b then
16                  swap the labels of t and b
17                  update hash tables using t and b
18                  not_moved =false
19      Until not_moved
20  End
```

**Fig. 1**. The LSA Algorithm.

In the initialization phase of the LSA algorithm, we firstly select the first *k* records from the data set to construct initial set of outliers. Each consequent record is labeled as non-outlier and hash tables for attributes are also constructed and updated.

In iteration phase, we read each record *t* that is labeled as non-outlier, its label is exchanged with each of the *k* outliers and the changes on entropy value are evaluated. If the entropy decreases, the point's non outlier label is exchanged with the outlier label of the point that achieved the best new value and the algorithm proceeds to the next object. After each swap, the hash tables are also updated. If no swap happened in one pass of all records, iteration phase terminates; otherwise, a new pass begins. Essentially, at each step we locally optimize the criterion. In this phase, the key step is computing the changed value of entropy. With the use of hashing technique, in $O(1)$ expected time, we can determine the frequency of an attribute value in corresponding hash table. Hence, we can determine the decreased entropy value in $O(m)$ expected time since the changed value is only dependent on the attribute values of two records to be swapped.

### 4.4 Time and Space Complexities

**Worst-case analysis:** The time and space complexities of the LSA algorithm depend on the size of dataset (*n*), the number of attributes (*m*), the size of every hash table, the number of outliers (*k*) and the iteration times (*I*).

To simplify the analysis, we will assume that every attribute has the same number of distinct attributes values, *p*. Then, in the worst case, in the initialization phase, the time complexity is $O(n*m*p)$. In the iteration phase, since the computation of value change on entropy requires at most $O(m*p)$ and hence this phase has time complexity $O(n*k*m*p*I)$. Totally, the algorithm has time complexity $O(n*k*m*p*I)$ in worst case.

The algorithm only needs to store *m* hash tables and the dataset in main memory, so the space complexity of our algorithm is $O((p+n)*m)$.

**Practical analysis:** Categorical attributes usually have *small* domains. Typical categorical attributes domains considered for clustering consist of less than a hundred or, rarely, a thousand attribute values. An important of implication of the compactness of categorical domains is that the parameter, *p*, can be regarded to be very small. And the use of hashing technique also reduces the impact of *p*, as discussed previously, we can determine the frequency of an attribute value in $O(1)$ expected time, So, in practice, the time complexity of LSA can be expected to be $O(n*k*m*I)$.

The above analysis shows that the time complexity of LSA is linear to the size of dataset, the number of attributes and the iteration times, which make this algorithm scalable.

### 4.5 Enhancement for Real Applications

The data sets in real-life applications are usually complex. They have not only categorical data but also numeric data. Sometimes, they are *incomplete*. In this section, we discuss the techniques for handling data with these characteristics in SOE1.

**Handling numeric data.** To process numeric data, we apply the widely used binning technique [38] and choose equal-width method for its feasibility in producing varied frequency values.

**Handling missing values** To handle incomplete data, we provide two choices. In the first choice, missing values in an incomplete object will not be considered when updating histograms. In the second choice, missing values are treated as special categorical attribute values. In our current implementation, we use the second choice.

## 5. Experimental Results

A comprehensive performance study has been conducted to evaluate our SOE1 algorithm. In this section, we describe those experiments and their results. We ran our algorithm on real-life datasets obtained from the UCI Machine Learning Repository [39] to test its performance against other algorithms on identifying true outliers. In addition, some large synthetic datasets are used to demonstrate the scalability of our algorithm.

### 5.1 Experiment Design and Evaluation Method

We used two real life datasets (*lymphography* and *cancer*) to demonstrate the effectiveness of our algorithm against *FindFPOF* algorithm [4], *FindCBLOF* algorithm [27] and *KNN* algorithm [16]. In addition, on the *cancer* dataset, we add the results of *RNN* based outlier detection algorithm [33,34] that are reported in [33,34] for comparison, although we didn't implement the *RNN* based outlier detection algorithm.

For all the experiments, the two parameters needed by *FindCBLOF* [27] algorithm are set to 90% and 5 separately as done in [27]. For the *KNN* algorithm [16], the results were obtained using the *5-nearest-neighbour*; For *FindFPOF* algorithm [4], the parameter *mini-support* for mining frequent patterns is fixed to 10%, and the maximal number of items in an itemset is set to 5. Since the LSA algorithm is parameter-free (besides the number of desired outliers), we don't need to set any parameters.

As pointed out by Aggarwal and Yu [3], one way to test how well the outlier detection algorithm worked is to run the method on the dataset and test the percentage of points which belong to the rare classes. If outlier detection works well, it is expected that the rare classes would be over-represented in the set of points found. These kinds of classes are also interesting from a practical perspective.

Since we know the true class of each object in the test dataset, we define objects in small classes as rare cases. The number of rare cases identified is utilized as the assessment basis for comparing our algorithm with other algorithms.

### 5.2 Results on Lymphography Data

The first dataset used is the Lymphography data set, which has 148 instances with 18 attributes. The data set contains a total of 4 classes. Classes 2 and 3 have the largest number of instances. The remained classes are regarded as rare class labels for they are small in size. The corresponding class distribution is illustrated in Table 1.

**Table 1.** Class Distribution of Lymphography Data Set

| Case | Class codes | Percentage of instances |
|---|---|---|
| Commonly Occurring Classes | 2, 3 | 95.9% |
| Rare Classes | 1, 4 | 4.1% |

Table 2 shows the results produced by different algorithms. Here, the *top ratio* is ratio of the number of records specified as *top-k* outliers to that of the records in the dataset. The *coverage* is ratio of the number of detected rare classes to that of the rare classes in the dataset. For example, we let LSA algorithm find the *top 7* outliers with the top ratio of 5%. By examining these 7points, we found that 6 of them belonged to the rare classes.

**Table 2:** Detected Rare Classes in Lymphography Dataset

| Top Ratio (Number of Records) | Number of Rare Classes Included (Coverage) | | | |
|---|---|---|---|---|
| | LSA | *FindFPOF* | *FindCBLOF* | *KNN* |
| 5% (7) | 6 (100%) | 5 (83%) | 4 (67%) | 4 (67%) |
| 10% (15) | 6 (100%) | 5 (83%) | 4 (67%) | 6 (100%) |
| 11% (16) | 6 (100%) | 6 (100%) | 4 (67%) | 6 (100%) |
| 15% (22) | 6 (100%) | 6 (100%) | 4 (67%) | 6 (100%) |
| 20% (30) | 6 (100%) | 6 (100%) | 6 (100%) | 6 (100%) |

In this experiment, the LSA algorithm performed the best for all cases and can find all the records in rare classes when the *top ratio* reached 5%. In contrast, for the *KNN* algorithm, it achieved this goal with the *top ratio* at 10%, which is almost the twice for that of our algorithm.

### 5.3 Results on Wisconsin Breast Cancer Data

The second dataset used is the Wisconsin breast cancer data set, which has 699 instances with 9 attributes, in this experiment, all attributes are considered as categorical. Each record is labeled as *benign* (458 or 65.5%) or *malignant* (241 or 34.5%). We follow the experimental technique of Harkins, et al. [33,34] by removing some of the *malignant* records to form a very unbalanced distribution; the resultant dataset had 39 (8%) *malignant* records and 444 (92%) *benign* records[1]. The corresponding class distribution is illustrated in Table 3.

**Table 3.** Class Distribution of Wisconsin breast cancer data set

| Case | Class codes | Percentage of instances |
|---|---|---|
| **Commonly Occurring Classes** | 1 | 92% |
| **Rare Classes** | 2 | 8% |

For this dataset, we also consider the *RNN* based outlier detection algorithm [33]. The results of *RNN* based outlier detection algorithm on this dataset are reproduced from [33].

Table 4 shows the results produced by the different algorithms. Clearly, among all of these algorithms, *RNN* performed the worst in most cases. In comparison to other algorithms, LSA always performed the best besides the case when top ratio is 4%. Hence, this experiment also demonstrates the superiority of LSA algorithm.

---
[1] The resultant dataset is public available at: http://research.cmis.csiro.au/rohanb/outliers/breast-cancer/

Table 4: Detected Malignant Records in Wisconsin Breast Cancer Dataset

| Top Ratio(Number of Records) | Number of Rare Classes Included (Coverage) | | | | |
|---|---|---|---|---|---|
| | LSA | FindFPOF | FindCBLOF | RNN | KNN |
| 1%(4) | **4 (10.26%)** | 3(7.69%) | **4 (10.26%)** | 3 (7.69%) | **4 (10.26%)** |
| 2%(8) | **8 (20.52%)** | 7 (17.95%) | 7 (17.95%) | 6 (15.38%) | **8 (20.52%)** |
| 4%(16) | 15(38.46%) | 14 (35.90%) | 14 (35.90%) | 11 (28.21%) | **16(41%)** |
| 6%(24) | **22(56.41%)** | 21 (53.85%) | 21 (53.85%) | 18 (46.15%) | 20(51.28%) |
| 8%(32) | **29(74.36%)** | 28(71.79%) | 27 (69.23%) | 25 (64.10%) | 27(69.23%) |
| 10%(40) | **33(84.62%)** | 31(79.49%) | 32 (82.05%) | 30 (76.92%) | 32(82.05%) |
| 12%(48) | **38 (97.44%)** | 35 (89.74%) | 35 (89.74%) | 35 (89.74%) | 37(94.87%) |
| 14%(56) | **39 (100%)** | **39 (100%)** | 38 (97.44%) | 36 (92.31%) | **39 (100%)** |
| 16%(64) | 39 (100%) | 39 (100%) | 39 (100%) | 36 (92.31%) | 39 (100%) |
| 18%(72) | 39 (100%) | 39 (100%) | 39 (100%) | 38 (97.44%) | 39 (100%) |
| 20%(80) | 39 (100%) | 39 (100%) | 39 (100%) | 38 (97.44%) | 39 (100%) |
| 25%(100) | 39 (100%) | 39 (100%) | 39 (100%) | 38 (97.44%) | 39 (100%) |
| 28%(112) | 39 (100%) | 39 (100%) | 39 (100%) | 39 (100%) | 39 (100%) |

### 5.4 Scalability Tests

The purpose of this experiment was to test the scalability of the LSA algorithm when handling very large datasets. A synthesized categorical dataset created with the software[2] developed by Dana Cristofor [40] is used. The data size (i.e., number of rows), the number of attributes and the number of classes are the major parameters in the synthesized categorical data generation, which were set to be 100,000, 10 and 10 separately. Moreover, we set the random generator seed to 5. We will refer to this synthesized dataset with name of DS1.

We tested two types of scalability of the LSA algorithm on large dataset. The first one is the scalability against the number of objects for a given number of outliers and the second is the scalability against the number of objects for a given number of objects. Our LSA algorithm was implemented in Java. All experiments were conducted on a Pentium4-2.4G machine with 512 M of RAM and running Windows 2000. Fig. 2 shows the results of using LSA to find 30 outliers with different number of objects. Fig. 3 shows the results of using LSA to find different number of outliers on DS1 dataset.

One important observation from these figures was that the run time of LSA algorithm tends to increase linearly as both the number of records and the number of outliers are increased, which verified our claim in Section 4.4.

---

[2] The source codes are public available at: http://www.cs.umb.edu/~dana/GAClust/index.html

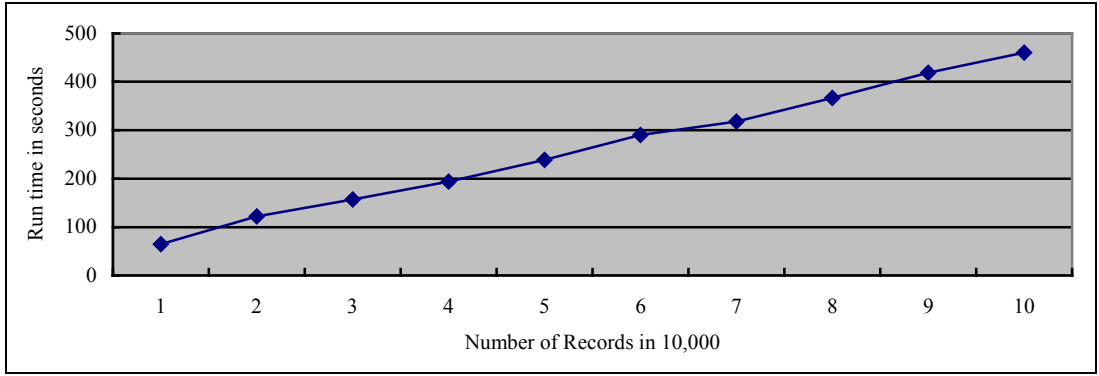

**Fig. 2**. Scalability of LSA to the number of objects when mining 30 outliers from DS1 dataset

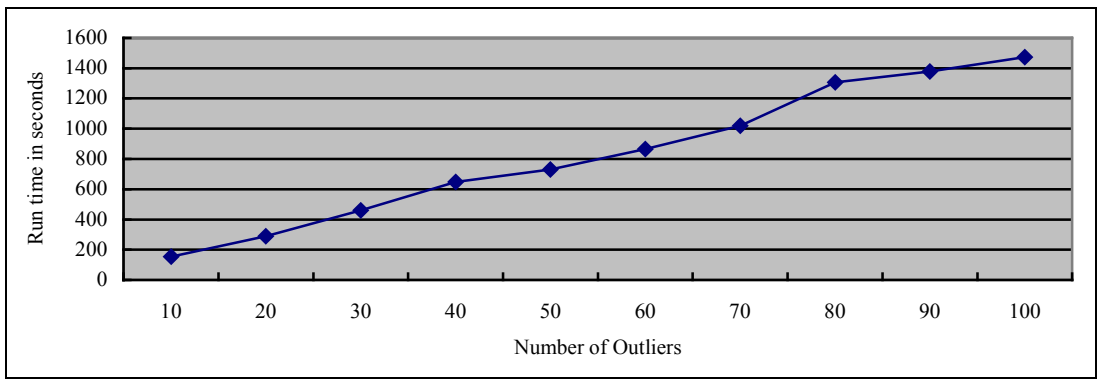

**Fig. 3**. Scalability of LSA to the number of outliers when mining outliers from 100,000 records of the DS1 dataset

## 6. Coclusions

The problem of outlier detection has traditionally been addressed using data mining methods. There are opportunities for optimization to improve these methods, and this paper focused on building an optimization model for outlier detection. Experimental results on real datasets and large synthetic datasets demonstrate the superiority of our new optimization-based method.

## Acknowledgements

This work was supported by The High Technology Research and Development Program of China (No. 2003AA4Z2170, No. 2003AA413021), the National Nature Science Foundation of China (No. 40301038) and the IBM SUR Research Fund.